# Stable Orbits in the Feeding Zone of the Planet Proxima Centauri *c*

## S. I. Ipatov*

*Vernadsky Institute of Geochemistry and Analytical Chemistry, Russian Academy of Sciences, Moscow, Russia *e-mail: siipatov@hotmail.com*

**Abstract**—Estimates of the size of the feeding zone of the planet Proxima Centauri *c* have been made at initial orbital eccentricities of planetesimals equal to 0.02 or 0.15. The research is based on the results of modeling of the evolution of planetesimals' orbits under the influence of the star and planets Proxima Centauri *c* and *b*. The considered time interval reached a billion years. It was found that after the accumulation of the planet Proxima Centauri *c* some planetesimals may have continued to move in stable elliptical orbits within its feeding zone, largely cleared of planetesimals. Usually such planetesimals can move in some resonances with the planet (Proxima Centauri *c*), for example, in the resonances 1:1 (as Jupiter Trojans), 5:4 and 3:4 and usually have small eccentricities. Some planetesimals that moved for a long time (1–2 million years) along chaotic orbits fell into the resonances 5:2 and 3:10 with the planet Proxima Centauri *c* and moved in them for at least tens of millions of years.

## INTRODUCTION

Bodies located in a certain neighborhood of the growing planet's orbit can fall onto it. This neighborhood is called the planet's feeding zone. Many papers have been published on estimates of the size of this zone and on the evolution of the orbits of two gravitationally interacting bodies moving around a massive center in close orbits (Artymowicz, 1987; Bruno, 1990; Gladman and Duncan, 1990; Goldreich and Tremaine, 1982; Gladman, 1993; Demidova and Shevchenko, 2016; Demidova and Shevchenko, 2020, 2021; Dermott, 1981a, 1981b; Farinella et al., 1994; Greenzweig and Lissauer, 1992; Duncan et al., 1989; Hasegawa and Nakazawa, 1990; Holt et al., 2020; Ida and Nakazawa, 1989; Ipatov, 1981, 1994, 2000; Kaplan and Cengiz, 2020; Kohne and Batygin, 2020; Levison et al., 1997; Lissauer and Kary, 1991; Markeev, 1978; Mikkola and Innanen, 1990, 1992; Morrison and Malhotra, 2015; Nishida, 1983; Petit and Henon, 1986; Rabe, 1961, 1962; Shevchenko, 2020a, 2020b; Szebehely, 1967; Tanikawa et al., 1991; Weiss-man and Wetherill, 1974; Zhang-yin and Lin, 1992; and others). Basically, such studies were carried out within the framework of the three-body problem (sun–planet–body), and the initially circular helio-centric orbits of the planet and body were considered more often.

The feeding zone of the planet is not monolithic. It cannot be said that all bodies with semimajor axes of the initial orbits that are in a certain interval will fall onto the planet or will be thrown into hyperbolic orbits after a sufficiently large time interval. Stable orbits within this interval, and unstable orbits outside this interval, are possible. For example, some bodies could remain inside the planet's feeding zone at the resonance 1:1 with the motion of the planet. Many Jupiter Trojans are known, but some bodies for this resonance have also been found for Earth, Mars, Uranus, and Neptune. At the same time, there are asteroid-free Kirkwood gaps.

In this paper, we first review the papers devoted to the study of the size of the feeding zones of the planets of the Solar System and the motion of bodies in stable orbits inside the feeding zones, as well as the estimation of the parameters of unstable orbits outside these zones. The following sections present the results of calculations for the planetary system Proxima Centauri. In this system, the mass of the star Alpha Centauri C (α Centauri C) is 0.1221 times the mass of the Sun. This star is a member of a triple star system, which also includes a binary star with masses of the order of the mass of the Sun: Alpha Centauri A (Rigil Kentaurus) and Alpha Centauri B (Toliman). The dis-tance between Alpha Centauri C (α Cen C) and Alpha Centauri AB (α Cen AB) is 12950 AU. Calculations performed in (Schwarz et al., 2018) showed that the gravitational influence of α Cen AB has little effect on the motion of exocomet bodies even in highly eccentric orbits around α Cen C.

The initial eccentric and inclined orbits of the planetesimals were in some vicinity of the orbit of the planet Proxima Centauri *c* (α Cen Cc). The gravita-tional infuence of the star and planets Proxima Centauri *c* and *b* (α Cen Cb) moving in eccentric orbits was taken into account. Below, when designating planets, the word Centauri is often omitted (for example, Proxima *c*). The main boundaries of the feeding zone of the planet Proxima *c* were estimated at different initial eccentricities of planetesimal orbits (equal to 0.02 or 0.15). The main attention was paid to the study of the orbits of planetesimals that remained inside the feeding zone of the planet Proxima *c*, and the initial values of the semimajor axes of the orbits of planetesimals that collided with planets or a star, or were thrown into hyperbolic orbits, although these values are outside the feeding zone.

## EVOLUTION OF THE ORBITS OF BODIES UNDER THE INFLUENCE OF A PLANET MOVING AROUND THE SUN

The study of gravitational interactions of two objects moving around a massive center has previously been carried out mainly for the Solar System. Several types of changes $N_t$ can be distinguished in the orbital elements of two gravitationally interacting bodies (Weissman and Wetherill, 1974; Gladman, 1993; Dermott, 1981a, 1981b; Duncan et al., 1989; Ipatov, 1981; 1994, 2000). Ipatov (1981, 1994, 2000) presented the results of studies of the interactions of two material points (MPs) moving around a massive central body, the Sun, in initially close orbits. Mutual gravitational infuence of MPs was taken into account by numerical integration of the equations of motion. In the case of initially circular orbits of two gravitationally interacting MPs, Ipatov (1994, 2000) considered the regions of the initial values of the semimajor axes of the MP orbits and the initial angle with the apex in the Sun between directions to the MPs, corresponding to several types of evolution. The range of values of the ratio μ (the sum of the masses of two MPs to the mass of the gravitating center (the Sun)) varied from $10^{-9}$ to $10^{-3}$. With the *N*-type, the graphs of changes in the semimajor axes of the MPs' orbits over time *t* have an *N*-shaped form. In this case, in synodic (rotating around the Sun with an angular velocity equal to the angular velocity of the first MP) coordinates, the orbit of the second MP covers one triangular libration point and, with an almost circular orbit of the first MP, looks like a tadpole (sickle). In the case of the *M*-type, the dependence of the semimajor axis *a* of the orbit on time *t* have an *M*-shaped form. In synodic coordinates, the orbit of the second MP covers both triangular libration points. In this case at initially circular sidereal orbits with $N_t = M$, the synodic orbit has a horseshoe shape. Close encounters of MPs are possible only for the *A*-type. Changes in the elements of the orbits in this case are chaotic and the semimajor axes of the orbits of the MPs at some moments of time can be the same. With the *C*-type, the elements of the orbits change chaotically, but there are no close encounters and the values of the semimajor axes of the MPs' orbits cannot become the same. If type $N_t = P$, then *a* and *e* change periodically, and the synodic orbit of the second MP envelops the Sun in the same way as with chaotic changes. For this type, a large number of subtypes can be distinguished, each of which is characterized by its own interrelationships of changes in orbital elements.

In the case of initially circular orbits at the initial angle $φ_0$ between directions from the Sun to MPs equal to 60° and $10^{-9} ≤ μ ≤ 2 × 10^{-4}$, the maximum values of $ε_0 = (a_{20} − a_{10})/a_{10}$, which correspond to the types *N*, *M*, *A* and *C*, were obtained equal to $α = (1.63 – 1.64)μ^{1/2}$, $β = (0.77 – 0.81)μ^{1/3}$, $γ = (2.1 – 2.45)μ^{1/3}$, and $δ = (1.45 – 1.64)μ^{2/7}$, respectively. Here $a_{10}$ and $a_{20}$ are the initial values of the semimajor axes *a* of orbits of two MPs, and μ is the ratio of the sum of MP masses to the solar mass. The γ values were close to the δ values. Values α, β, γ and δ are generally smaller at other (than 60°) values $φ_0$. For $ε_0=0$, the smallest values of $φ_0$, which correspond to the types *N* and *M*, are approximately equal to 0.4 and $4μ^{1/3}$ radians, respectively. In the case of a large difference in the masses of objects (planets and bodies) that initially moved in circular orbits, the maximum eccentricities of the bodies' orbits at $μ≤10^{-5}$ usually did not exceed $(7–8)μ^{1/3}$ for type *A* and $(4–6)μ^{1/3}$ for type *C*.

In this paragraph, for the case of initially circular orbits, we present the values of α, β, γ and δ, obtained in other papers. Ida and Nakazawa (1989) found $β = 1.24μ^{1/3}$ analytically and $β = 1.3μ^{1/3}$ with numerical estimates. Nishida (1983) found $β = 1.04μ^{1/3}$. In numerical calculations, Gladman and Duncan (1990) and Gladman (1993) obtained $γ = 2.1μ^{1/3}$ at $φ_0 = 0$ and $γ = 2.4μ^{1/3}$ at $φ_0 = 180°$. Exploring the Jacobi integral in the restricted three-body problem, Birn (1973) showed that $γ ≈ 2.4 \ μ^{1/3}$. Demidova and Shevchenko (2020) considered $1.38a_μ^{0.29}$ and $2.54a_μ^{0.34}$ for the inner and outer boundaries of the chaotic zone, respectively. The values of the boundaries of the regions of initial distances from the gravitating center (in the radii of the Hill sphere), corresponding to various final eccentricities, are shown in Fig. 3 in (Demidova and Shevchenko, 2021). These values are slightly smaller than the values of β and γ given above. Duncan etc. (1989) numerically obtained $δ = 1.49μ^{2/7}$, and in (Morrison and Malhotra, 2015) $δ = 1.5μ^{2/7}$, in (Shevchenko, 2020a) $δ = 1.62μ^{2/7}$. Gladman (1993) studied the values of γ for the same body masses. Other authors cited in this paragraph considered only the case of zero mass of the second body. Ranges of $ε_0$ values, at which during the time $t < T_S$ ($T_S$ is the synodic period of rotation), a collision of objects occurs, were studied by a number of authors (Greenzweig and Lissauer, 1992; Ida and Nakazawa, 1989; Lissauer and Kary, 1991; Petit and Henon, 1986; Tanikawa et al., 1991) in the study of the formation of axial rotation and planetary accretion rates. Numerical studies in (Lissauer and Kary, 1991) showed that most planetesimals at $e = i = 0$ ($e$ and $i$ are the eccentricities and inclinations of the planetesimal orbits) and $0.9\mu^{1/3} < \varepsilon_0 < 2.2\mu^{1/3}$ approaches the planet at a distance of $0.1h$ ($h = a\,(\mu/3)^{1/3}$) for $20T_s$.

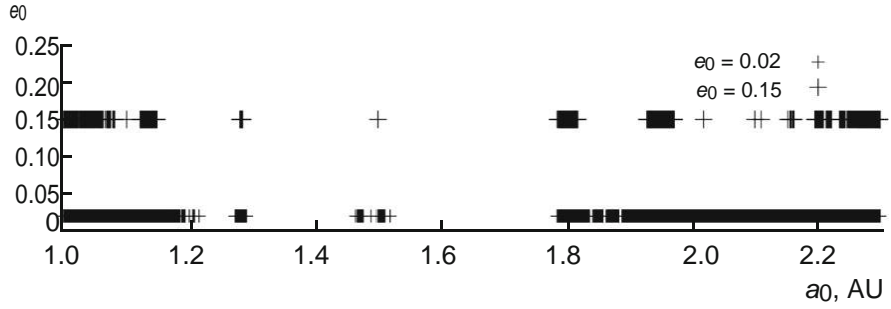

**Fig. 1.** Initial values $a_0$ of the semimajor axes of the orbits (in AU) of the planetesimals, which had elliptical orbits at the end of the considered time interval at $e_0 = 0.02$ and $e_0 = 0.15$. The semimajor axis of the orbit of the planet Proxima Centauri $c$ is equal to 1.489 AU.

Calculations carried out in (Mikkola and Innanen, 1990) showed that $\beta$ decreases with increasing $i$. When studying the circular problem of three bodies in the case of $\mu \ll 1$, $e \ll 1$, $i \ll 1$ rad and $a \approx$ const, it was found (Artymowicz, 1987; Ida and Nakazawa, 1989; Hasegawa and Nakazawa, 1990) that $\gamma^2 = 4(e^2 + i^2)/3 + 12(h/a)^2$. Greenzweig and Lissauer (1992) and Tanikawa et al. (1991) also obtained an increase in $\gamma$ and a decrease in $\beta$ with increasing $e$. Gratia and Lissauer (2020) studied times elapsed before close encounters in a system of five Earth-mass planets as a function of the initial distances between the planets' orbits. In (Goldberg et al., 2022), criteria for the stability of planetary orbits in a chain of resonances were considered.

Motion of bodies at the resonance 1:1 (including near triangular libration points) was considered, for example, in (Goldreich and Tremaine, 1982; Der-mott, 1981a, 1981b; Ipatov, 1994, 2000; Levison et al., 1997; Markeev, 1978; Mikkola and Innanen, 1990; 1992; Qi and Qiao, 2022; Rabe, 1961, 1962; Szebehely, 1967; Weissman and Wetherill 1974; Zhang-yin and Lin, 1992). As a larger body, specific bodies of the Solar System were taken mainly. In (Dermott, 1981a) it was analytically shown that $\alpha$ is proportional to $\mu^{1/2}$. Triangular libration points are unstable in the sense of Lyapunov for $\mu > (9 - \sqrt{69})/18 \approx 0.04$ (Markeev, 1978). According to (Jewitt et al., 2000), the mean orbital inclination of the Jupiter Trojans is 13.7°. It was noted in (Holt et al., 2020) that the orbits of some Jupiter Trojans are stable over the lifetime of the Solar System. In (Kohne and Batygin, 2020), the relation-ship between Jupiter's reverse Trojans and highly inclined centaurs and trans-Neptunian objects was considered. Kaplan and Cengiz (2020) studied the motion of small bodies in the resonance 1 : 1 with the

Earth (eleven bodies in horseshoe synodic orbits and one Trojan).

In some cases (especially for $\mu_1 \sim 10^{-5} \gg \mu_2$, where $\mu_1$ and $\mu_2$ are the ratios of the masses of the first and second MPs to the mass of the gravitating center) for $N_t = A$, Ipatov (1994, 2000) obtained transfers of MPs to resonant orbits. Usually, after several hundred revolutions of the MPs around the Sun, these resonant relationships were violated. For most of the considered resonances ($T_1 : T_2 = 1 : 2, 5 : 12, 5 : 13, 5 : 6, 5 : 7, 5 : 4$, $T_i$ is the period of revolution of the $i$th MP around the Sun), the semimajor axes and eccentricities of the orbits changed periodically with a small amplitude, and the changes in the longitude of the perihelion were small. Motion around resonances in these cases are oscillations around periodic solutions representing closed curves in synodic coordinates and have been studied by many authors (see, for example, Bruno, 1990, and chapters 8 and 9 in (Szebehely, 1967)). In one of the calculation options (Fig. 2.4 in (Ipatov, 2000)) at the resonances 5:6 and 4:5, the semimajor axis of the orbit of the smaller body changed insignificantly, the eccentricity increased almost monotonically, and the longitude of the perihelion decreased. In these cases, in synodic coordinates, the curve around which small oscillations occurred was not closed. In their Fig. 2, Demidova and Shevchenko (2020) considered the distribution of particles in the disk with distance from the star after $10^4$ revolutions of the planet around the star for a number of values of the masses of the planet. This figure shows a decrease in the number of particles at resonances $2 : 1, 3 : 2, 4 : 3, 5 : 4, 4 : 5, 3 : 4$ and $1 : 2$ and an increase in the number of particles at the resonance $1 : 1$. For the resonances $2 : 1$ and $1 : 2$, a similar decrease was noted in (Demidova and Shevchenko, 2016).

In (Farinella et al., 1994; Ipatov, 1989, 1992a, 1992b; Kazantsev and Kazantseva, 2021; Scholl and Froeschle, 1975, 1990, 1991; Sidlichovsky and Melendo, 1986; Wisdom, 1982, 1983; Yoshikawa, 1991 etc.) it is shown that asteroids that were in some resonances with Jupiter (3 : 1, 5 : 2, 2 : 1, 7 : 3) and in secular resonances ($\nu_5$, $\nu_6$, $\nu_{16}$), could significantly increase the eccentricities of their orbits and leave resonances due to approaches to the planets. The dependences of inclinations *i* on the semimajor axes *a* of orbits for secular resonances are shown in Fig. 8.21 in the monograph (Shevchenko, 2020b).

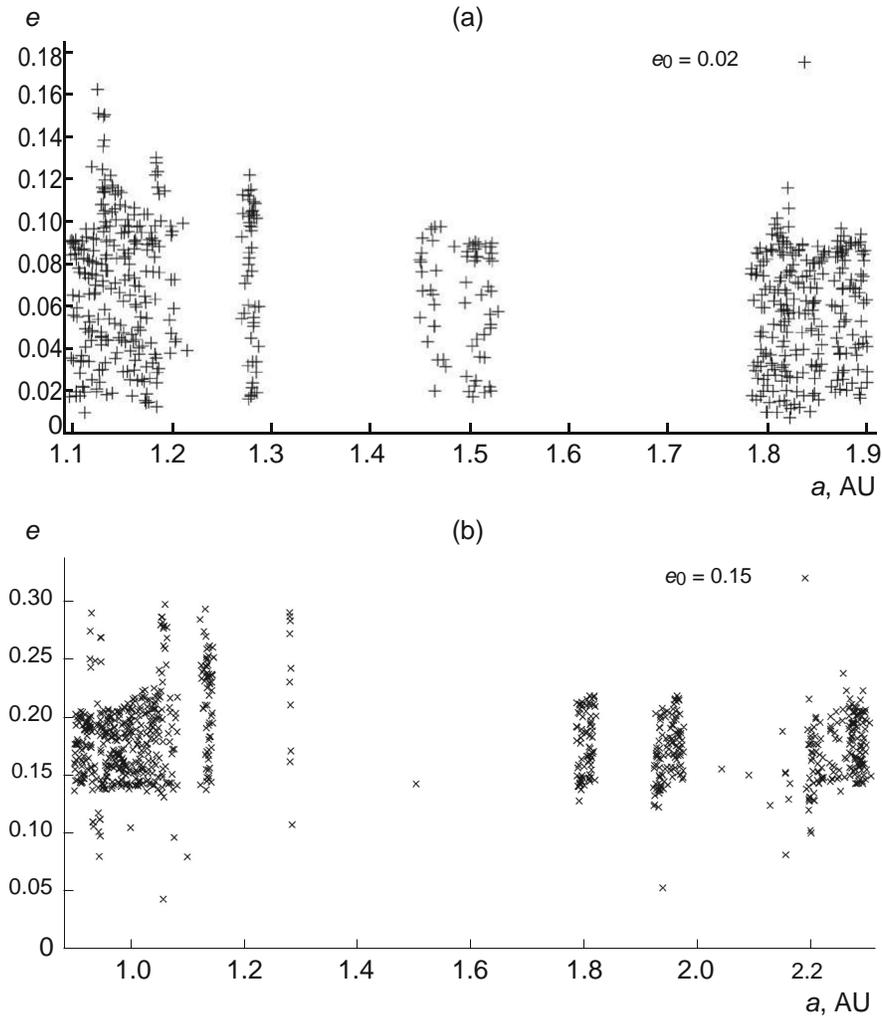

**Fig. 2.** Values of eccentricities *e* of orbits depending on the values *a* of semimajor axes of the orbits (in AU) of the planetesimals at the final considered moments of time at $e_0 = 0.02$ (Fig. 2a) and $e_0 = 0.15$ (Fig. 2b). The semimajor axis of the orbit of the planet Proxima Centauri *c* is equal to 1.489 AU.

Shevchenko (2022) considered the zone of weak chaos between the resonances 2 : 1 and 1 : 1 with the planet. He showed that for Jupiter the times of removal of bodies from this zone did not generally exceed $10^9$ years. For the Earth, these times were longer. In this paper, it is shown that the material captured in the resonances 3:2 and 4:3 with Jupiter, could survive. Asteroids of the Hilda and Thule families are examples of such resonances. According to (Dvorak and Kubala, 2022), the semimajor axes of relatively stable orbits are in the range from 7.12 to 7.23 AU. This interval is near the 5:8 (7.12 AU) and 3 : 2 (7.28 AU) resonances with Jupiter.

## VARIANTS OF CALCULATIONS

The following estimates of the parameters of the feeding zone of Proxima Centauri *c* (α Cen Cc) are based on the results of calculations of the evolution of the orbits of planetesimals, which were originally located in some neighborhood of the orbit of this planet, located beyond the ice line. In (Ipatov, 2021, 2022) it was found that the mass of water delivered from the feeding zone of this planet to the inner planet Proxima Centauri *b* (α Cen Cb) could exceed the mass of the Earth's oceans, and a slightly smaller number of icy planetesimals could be delivered to the planet Proxima Centauri *d* (α Cen Cd). The number of planetesimals thrown into hyperbolic orbits was no less than the number of planetesimals that collided with the planets. Below we discuss the size of the feeding zone of the planet Proxima *c*, and the orbits of individual planetesimals remaining within this feeding zone.

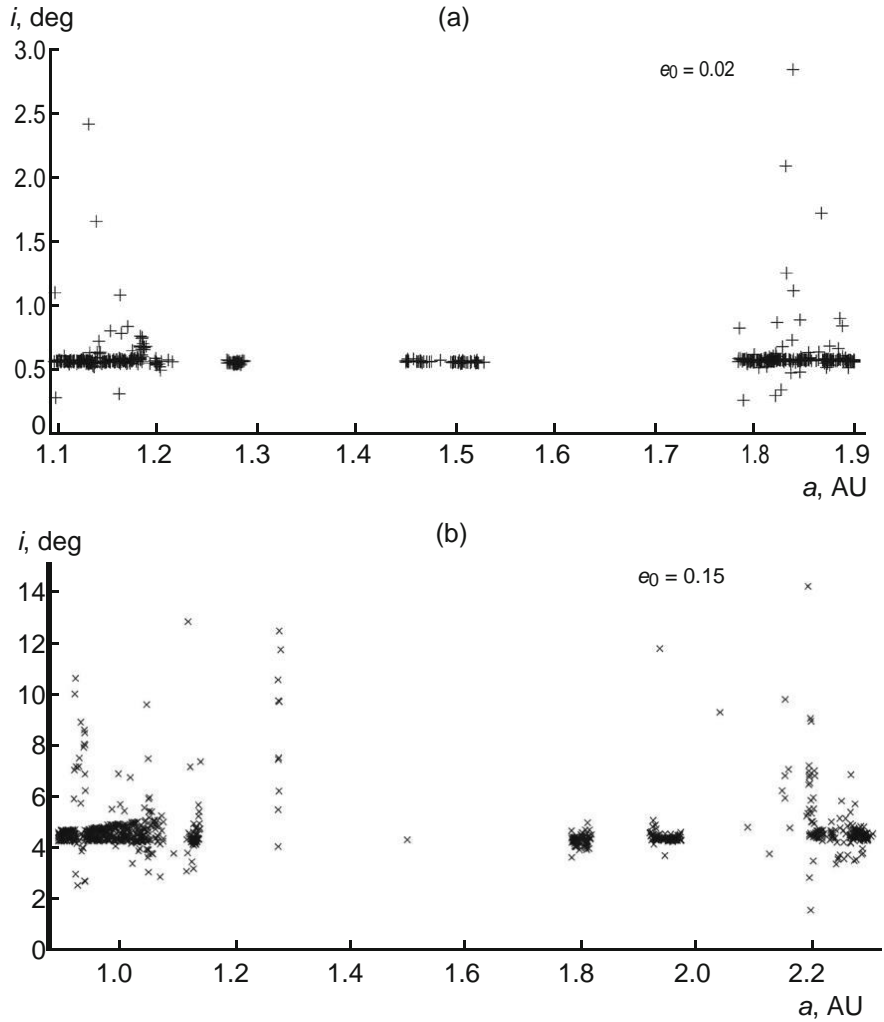

**Fig. 3.** Values of inclinations $i$ of orbits (in degrees) depending on the values $a$ of semimajor axes of the orbits (in AU) of planetesimals at the final considered moments of time at $e_0 = 0.02$ (Fig. 3a) and $e_0 = 0.15$ (Fig. 3b). The semimajor axis of the orbit of the planet Proxima Centauri $c$ is equal to 1.489 AU.

When modeling the motion of planetesimals, the gravitational influence of the star with a mass equal to 0.1221 solar masses, the planet Proxima $c$, ($a_c = 1.489$ AU, $e_c = 0.04$, $m_c = 7m_E$, $m_E$ is the mass of the Earth) and the planet $b$ ($a_b = 0.04857$ AU, $e_b = 0.11$, $m_b = 1.17m_E$) were taken into account. Calculations were also carried out with the mass of the embryo of Proxima $c$ equal to $k_c = 0.5$ or $k_c = 0.1$ of its present mass $m_c$. Unless otherwise stated, below we are talking about calculations with $k_c = 1$ ($m_c = 7m_E$). The orbital inclinations of the planets were taken equal to zero. The initial eccentricities of the planetesimal orbits were equal to $e_0 = 0.02$ or $e_0 = 0.15$. Initial eccentricities of orbits of planetesimals were taken small, and their large eccentricities could be formed as a result of the mutual gravitational influence of the planetesimals. Calculations of the evolution of the disks of bodies corresponding to the feeding zone of the terrestrial planets showed an increase in the average eccentricity of the orbits of the bodies up to 0.2 and even up to 0.4 at the last stages of disk evolution (Ipatov, 1993, 2000). The initial inclinations of the planetesimal orbits were equal to $e_0/2$ radians (i.e., equal to 0.57° or 4.3° at $e_0 = 0.02$ or $e_0 = 0.15$, respectively). In each calculation variant, the initial values of semimajor axes $a_0$ of planetesimal orbits ranged from $a_{min}$ to $a_{max} = a_{min} + 0.1$ AU. For the ($i + 1$)th planetesimal, the value of $a_0$ was calculated by the formula $a_0(i + 1) = (a_0^2 i + [(a_{min} + d_a)^2 - a_{min}^2 ] / N_0 )^{1/2}$, where $a_{0i}$ is the value of $a_0$ for the $i$th planetesimal, $d_a = 0.1$ AU, and the initial number of planetesimals in each variant (with fixed values of $a_{min}$ and $e_0$) was equal to $N_0 = 250$. Values of $a_{min}$ varied from 0.9 to 2.2 AU with a step of 0.1 AU. The considered time interval in the calculations was not less than 100 million years (if the evolution did not end earlier). In variants with $a_{min}$ from 1.2 to 1.7 AU, the calculations were carried out over several hundred million years (up to 1000 million years). In contrast to the problem of three bodies with circular initial orbits, considered in (Ipatov, 1994; 2000), in the considered studies of the motion of planetesimals in the Proxima Centauri system, the initial eccentricities of the orbits of the planet Proxima $c$ ($e_{c0}$ = 0.04) and planetesimals ($e_0$) were not equal to zero.

The equations of motion were integrated using the symplectic algorithm from the SWIFT package (Levison and Duncan, 1994). Bodies that collided with planets or a star or reached 1200 AU were excluded from the integration. According to (Schwarz et al., 2018), the radius of the Hill sphere of the star Proxima Centauri (α Cen C) is 1200 AU. Calculations with different integration time steps $t_s$, equal to 0.1, 0.2, 0.5, 1, or 2 Earth days, gave approximately the same results (taking into account the fact that the evolution of orbits is chaotic at close encounters). Below are the results for $t_s$ equal to one day (except for Figs. 10, 11). It was noted in (Frantseva et al., 2022) that in the algorithm (Levison and Duncan, 1994) the integration step decreases significantly at distances smaller than 3.5 Hill radii.

## FEEDING ZONE OF THE EXOPLANET PROXIMA CENTAURI $c$

Planet Proxima Centauri $c$ moves far from the planets Proxima Centauri $b$ and $d$ and has a much larger mass than these two planets. Therefore, the motion of most planetesimals in orbits close to the orbit of the planet Proxima $c$ can be close to the motion of bodies for the three-body problem. Initial values $a_0$ of semimajor axes of the orbits of planetesimals, which still had elliptical orbits at the end of the considered evolution at $e_0$ = 0.02 and $e_0$ = 0.15, are shown in Fig. 1. Basic boundaries between values of $a_{i0}$, corresponding to planetesimals thrown into hyperbolic orbits or colliding with planets, and planetesimals still moving in elliptical orbits, were equal to $a_{min002}$ = 1.194 AU and $a_{max002}$ = 1.786 AU at $e_0$ = 0.02 and were considered to be equal to $a_{min015}$ = 1.082 AU and $a_{max015}$ = 2.238 AU at $e_0$ = 0.15. For $e_0$ = 0.15 and 1.786 ≤ $a_{i0}$ ≤ 2.237 AU for some $a_{i0}$ planetesimals remained in elliptical orbits, but at other close values of $a_{i0}$ planetesimals were thrown into hyperbolic orbits.

For the planet Proxima $c$ with mass $m_c = 7m_E$ we have $\mu$ = 1.721 × $10^{-4}$, since the ratio of the mass of the star to the mass of the Sun is 0.1221. This value of $\mu$ corresponds to $57m_E$ in our Solar System. For Proxima $c$ we have $\mu^{1/3}$ = 0.0556, $(\mu/3)^{1/3}$ = 0.0386, $a_c\mu^{1/3}$ = 0.0828 AU, and the radius of the Hill sphere is $a_c(\mu/3)^{1/3}$ = 0.0574 AU. The calculation results showed that

$$a_c - a_{min002} = 3.5a_c \cdot \mu^{1/3} = 0.295 \text{ AU},$$
$$a_{max002} - a_c = 3.6a_c \cdot \mu^{1/3} = 0.298 \text{ AU},$$
$$a_c - a_{min015} = 4.9a_c \cdot \mu^{1/3} = 0.407 \text{ AU},$$
$$a_{max015} - a_c = 9.0a_c \cdot \mu^{1/3} = 0.749 \text{ AU}$$

If we consider the initial values of the product $a \times e$, which characterizes changes in the distance from the star to the moving planets and planetesimals (for example, $0.04a_c$ and $0.02a_{min002}$), then we have

$$a_c - a_{min002} = 0.04a_c + 0.02a_{min002} + 2.54a_c \cdot \mu^{1/3},$$
$$a_{max002} - a_c = 0.04a_c + 0.02a_{max002} + 2.40a_c \cdot \mu^{1/3},$$
$$a_c - a_{min015} = 0.04a_c + 0.15a_{min015} + 2.23a_c \cdot \mu^{1/3}, \text{ and}$$
$$a_{max015} - a_c = 0.04a_c + 0.15a_{max015} + 4.3a_c \cdot \mu^{1/3}$$

($e_c$ = 0.04 and $e_0$ = 0.02 or $e_0$ = 0.15). Coefficients before $a_c\mu^{1/3}$ in the above three formulas are approximately 2.2–2.5, i.e., are close to the coefficients in γ = $(2.1–2.45)\mu^{1/3}$ for circular initial orbits. For $a_{max015} - a_c$ the coefficient 4.3 was higher than the above values. For the latter case, the coefficient 2.5 corresponds to $a_{max015}$ = 2.07 AU. For $a_{i0}$ between 1.79 and 2.237 AU and $e_0$ = 0.15, there were a few planetesimals that were still moving in elliptical orbits at the end of the considered time, but most of the planetesimals for such an interval of $a_{i0}$ were thrown into hyperbolic orbits or collided with the planet Proxima $c$. The above studies may be of interest for studying the zones of initial values of the semimajor axes of the orbits of bodies that could continue moving along elliptical orbits in some other planetary systems with one dominant planet.

With the mass of the embryo of the planet Proxima $c$ equal to $k_c$ = 0.1 of its present mass $m_c$, (i.e., with a mass equal to $0.7m_E$) and $e_0$ = 0.02, it was found that $a_c\mu^{1/3}$ = 0.0266 AU, $a_{max002}$ = 1.5953 AU, $a_{max002} - a_c$ = $4.0a_c\mu^{1/3} = 0.04a_c + 0.02a_{max002} + 0.54a_c\mu^{1/3}$ = 0.1063 AU. Here the coefficient 0.54 before $a_c\mu^{1/3}$ is less than the coefficient 2.4 at $k_c$ = 1.

## MOTION OF PLANETESIMALS IN STABLE ORBITS WITHIN THE FEEDING ZONE OF THE PLANET PROXIMA CENTAURI $c$

Figure 1 shows that inside the intervals ($a_{min002}$, $a_{max002}$) and ($a_{min015}$, $a_{max015}$), which estimate the size of the feeding zone of the planet Proxima Centauri $c$, there are some initial values $a_{i0}$ of the semimajor axes of planetesimal orbits corresponding to planetesimals continuing to move in elliptical orbits, and there are some values of $a_{i0}$, corresponding to planetesimals thrown into hyperbolic orbits or colliding with planets outside such intervals. Such subregions usually correspond to resonances of the mean motions of planetesimals with the planet Proxima $c$. Examples of such subregions are presented in Tables 1 and 2. Some small subregions were not included in the tables. In Fig. 2, for values $a$ of semimajor axes of the orbits of planetesimals at the final considered moments of time (not for the initial values $a_{i0}$, as for Fig. 1 and Tables 1 and 2), the eccentricities of the orbits at these times are shown. Figure 3 shows the inclinations of the orbits (in degrees) for these values of $a$ and the final times. Figure 3a did not include the point ($a = 1.147$ AU, $i = 4.6°$). In Fig. 2, almost all values of eccentricities, including those for the subregions given in Table 1 did not exceed 0.15 and 0.3 at $e_0 = 0.02$ and $e_0 = 0.15$, respectively. In Fig. 3, most values of inclinations $i$ of orbits did not exceed 1° and 10° at $e_0 = 0.02$ and $e_0 = 0.15$, respectively. A significant part of the $i$ values differed from the initial values by no more than 0.2° and 1°, respectively, at $e_0 = 0.02$ and $e_0 = 0.15$.

Orbital eccentricities greater than 0.14 in Fig. 2 at $e_0 = 0.02$ were obtained for planetesimals moving in the resonance 3 : 2 with the motion of the planet Prox-ima $c$. Orbital eccentricities greater than 0.25 in Fig. 2 at $e_0 = 0.15$ were obtained for planetesimals moving in the 2 : 1, 5 : 3, 4 : 3 and 5 : 4 resonances with the motion of Proxima $c$.

In Fig. 2, orbital eccentricities of planetesimals that were in the resonance 1:1 with the planet Proxima $c$, did not exceed 0.1 at $e_0 = 0.02$ and were equal to 0.15 at $e_0 = 0.15$. Note for comparison that in Fig. 1 in the paper (Holt et al., 2020) it is seen that the orbital eccentricities of the Jupiter Trojans do not exceed 0.2. For Jupiter Trojans, almost all inclinations were less than 35°, and their mean inclination was 13.7° (Jewitt et al., 2000). Orbital inclinations of planetesimals at the resonance 1:1 in Fig. 3 were close to the initial inclinations in the calculations, that is, they were generally several times smaller than for the Jupiter Trojans.

The orbits of two known Earth Trojans have eccentricities of 0.19 and 0.38 and inclinations of 20.9° and 13.8° (https://en.wikipedia.org/wiki/2010_TK7, https://en.wikipedia.org/wiki/(614689)_2020_XL5). More asteroids in the resonance 1:1 with the motion of the Earth, moves in synodic coordinates in horseshoe orbits, and not around one of the libration points. The values of their eccentricities and inclinations can reach 0.5° and 20°, respectively (Kaplan and Cengiz, 2020). Possibly larger than in Fig. 3, orbital inclinations of asteroids in the resonance 1:1 with the motion of Jupiter or the Earth are related to the fact that in our calculations the orbits of planetesimals from the vicinity of the planet Proxima $c$ were swayed mainly only by this planet.

Examples of mean motion resonances for stable orbits include the resonance 5 : 4 for both values $e_0$, as well as the mean-motion resonances 3 : 2, 3 : 4, 2 : 3, 4 : 7 and 5 : 9 for $e_0 = 0.15$. At $e_0 = 0.02$ near the resonance 1:1 there were two subregions with the ratio $n_{rel}$ of the mean motion of planetesimals to the mean motion of the planet Proxima $c$ close to 0.98 and 1.02. Only one planetesimal (with $a_{i0} ≈ 1.5004$ AU) was still moving near the resonance 1:1 for $e_0 = 0.15$ at $T = 1000$ Myr. The value of $n_{rel} = 1.42$ in Table 1 is close to 7/5. Asteroid 279 Thule in the Solar System moves in the resonance 4:3 with Jupiter. In our calculations for the planet Proxima $c$ there were no planetesimals in the resonance 4:3 after a few hundred million years.

**Table 1.** Subregions ($a_{min i}$, $a_{max i}$) of initial values $a_{i0}$ (in AU) of semimajor axes of the orbits for which planetesimals continued to move in elliptical orbits at the end of the considered time interval. $n_{rel}$ is the ratio of the mean motion of planetesimals to the mean motion of the planet Proxima Centauri $c$

| $e_0$ | $a_{min i}$, $a_{max i}$ (AU) | $n_{rel}$ |
|---|---|---|
| 0.02 | 1.205,  1.209 | 1.42 |
| 0.02 | 1.273,  1.291 | 5/4 |
| 0.02 | 1.467,  1.476 | 1.02 |
| 0.02 | 1.500,  1.512 | 0.98 |
| 0.15 | 1.125,  1.149 | 3/2 |
| 0.15 | 1.282,  1.284 | 5/4 |
| 0.15 | 1.789, 1.818 | 3/4 |
| 0.15 | 1.928, 1.972 | 2/3 |
| 0.15 | 2.155, 2.159 | 4/7 |

**Table 2.** Subregions ($a_{min i}$, $a_{max i}$) of initial values $a_{i0}$ (in AU) of semimajor axes of orbits for which planetesimals were ejected into hyperbolic orbits or collided with planets. $n_{rel}$ is the ratio of the mean motion of planetesimals to the mean motion of the planet Proxima Centauri $c$

| $e_0$ | $a_{min i}$, $a_{max i}$ (AU) | $n_{rel}$ |
|---|---|---|
| 0.02 | 1.186,  1.187 | 7/5 |
| 0.02 | 1.838,  1.841 | 8/11 |
| 0.02 | 1.860,  1.862 | 4/5 |
| 0.02 | 1.884, 1.888 | 7/10 |
| 0.15 | 1.024,  1.025 | 7/4 |
| 0.15 | 1.052,  1.053 | 1.68 |
| 0.15 | 1.064,  1.068 | 1.65 |
| 0.15 | 2.208, 2.213 | 5/9 |
| 0.15 | 2.223, 2.234 | 0.546 |

Mean motion resonances for $a_{i0}$, in which planetesimals were ejected into hyperbolic orbits or collided with planets, included the resonances 5 : 4, 8 : 11, 4 : 5 and 7 : 10 for $e_0 = 0.02$ and the resonance 7 : 4 for $e_0 = 0.15$. In Table 2, values of the ratio $n_{rel}$ of the mean motion of planetesimals to the mean motion of the planet Proxima $c$, equal to 1.68 and 1.65, are close to 5/3. For this resonance in Fig. 1, instead of one gap, there are two closely spaced narrow gaps.

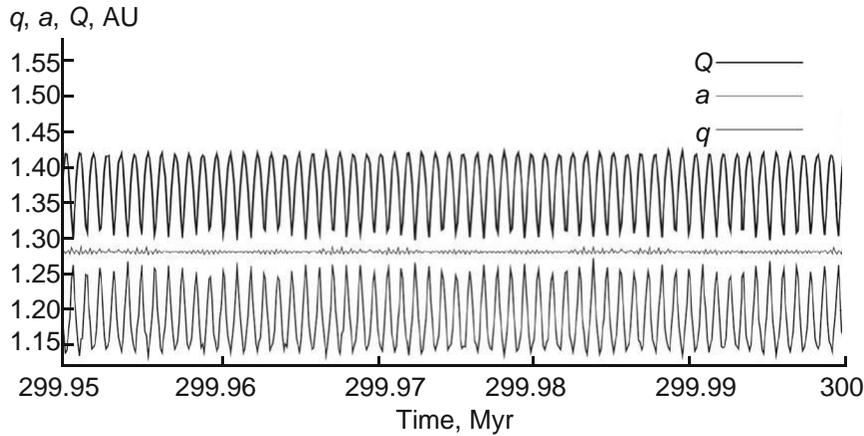

**Fig. 4.** Evolution with time (in Myr) of the semimajor axis $a$, perihelion and aphelion distances $q$ and $Q$ (in AU) of the orbits of the planetesimal at $a_0 = 1.28062$ AU, $a_{min} = 1.2$ AU, $e_0 = 0.02$, $k_C = 1$, and $t_S = 1^d$. The planetesimal moved in the 5:4 resonance with the planet Proxima Centauri $c$.

In planetary systems with one dominant planet, such as Proxima Centauri, within the planet's feeding zone there could be several subregions (not only Trojans) of semimajor axes of orbits in which bodies can move for a long time. Analogs of the asteroid and trans-Neptunian belts near Proxima Centauri (α Cen C) may be larger than that of the Solar System. The smaller ratio of the mass of the planet Proxima $c$ to the mass of the star than that of Jupiter, the greater ratio of the semimajor axes of the orbits of the planets of Proxima Centauri $c$ and $b$ than a similar ratio for Jupiter and Mars, and only one large planet in the Proxima Centauri system can be the reasons for such possible differences in the belts and the possible existence of a planet(s) between the orbits of the Proxima planets $b$ and $c$. Collisions and mutual gravitational influence of planetesimals could prevent long-term motion of planetesimals along stable orbits inside the planet's feeding zone. The same effects could contribute to the entry of bodies into subregions outside the planet's feeding zone, from which bodies can be removed due to their motion at certain resonances.

In Figs. 4–5 there are examples of changes of $a$, $q = a(1 - e)$, and $Q = a(1 + e)$ over time for the orbits of planetesimals moving in the 5 : 4 resonance with the planet Proxima $c$ at $e_0 = 0.02$ and $e_0 = 0.15$. These planetesimals were still moving in this resonance, although the figures are given for 300 or 910 million years after the beginning of the evolution of the orbits. Elements $e$, $q$ and $Q$ of their orbits changed with the period $T_0$, equal to 885 years with $e_0 = 0.02$ and 36 thousand years at $e_0 = 0.15$, i.e., the period was much longer for large initial eccentricities. In Figs. 6–7, the changes of $a$, $q$ and $Q$ are shown over time for orbits of planetesimals moving in the 1:1 resonance with the planet Proxima $c$. The periods of the major changes in

$q$ and $Q$ are about 15 thousand years with $e_0 = 0.02$ and 8 thousand years at $e_0 = 0.15$. These changes basically correspond to the changes in the semimajor axis in Fig. 6 and the eccentricity changes in Fig. 7. The planetesimals were still moving in the resonance 1:1 for hundreds of millions of years later. In Figs. 8–9, the planetesimals moved in the resonance 3:4. For these figures, the period $T_q$ is three thousand years at $e_0 = 0.02$ and 12.45 thousand years at $e_0 = 0.15$. The ratio of periods $T_q$ at $e_0 = 0.15$ to periods at $e_0 = 0.02$ was 4 and 41 for the resonances 3:4 and 5:4, respectively. The maximum eccentricities of planetesimal orbits in the variants shown in Figs. 4, 6 and 8 (with $e_0 = 0.02$) were 0.12, 0.09, and 0.12, respectively. In the variants shown in Figs. 5, 7 and 9 (with $e_0 = 0.15$), such maximum eccentricities were 0.31, 0.15, and 0.22, respectively.

Figures 10 and 11 show examples of the evolution of orbits in which a planetesimal moving for a long time (1–2 Myr) along a chaotic orbit fell into the resonances 5:2 and 3:10 with the planet Proxima $c$. In the variants presented in these figures, the planetesimals continued to move in these resonances throughout the considered time interval (equal to 100 million years).

In calculations with the mass of the embryo of the planet Proxima $c$ equal to $k_C = 0.5$ or $k_C = 0.1$ of its present mass $m_C$ (i.e., equal to $3.5m_E$ or $0.7m_E$), at $e_0 = 0.02$ after 100 Myr, the planetesimals continued to move in elliptical orbits in the case when the initial semimajor axes $a_{i0}$ (in AU) orbits were within (1.464, 1.514) and (1.472, 1.507) at $k_C = 0.5$ and $k_C = 0.1$, respectively. At $k_C = 1$ and $e_0 = 0.02$ in the middle of a similar interval for stable orbits, which is located near the resonance 1 : 1 with the planet Proxima $c$, there was a subinterval (1.476, 1.500 AU) corresponding to planetesimals ejected into hyperbolic orbits or colliding with planets (see Table 1). At $e_0 = 0.15$, the interval of $a_{i0}$ (in AU), corresponding to the planetesimals that

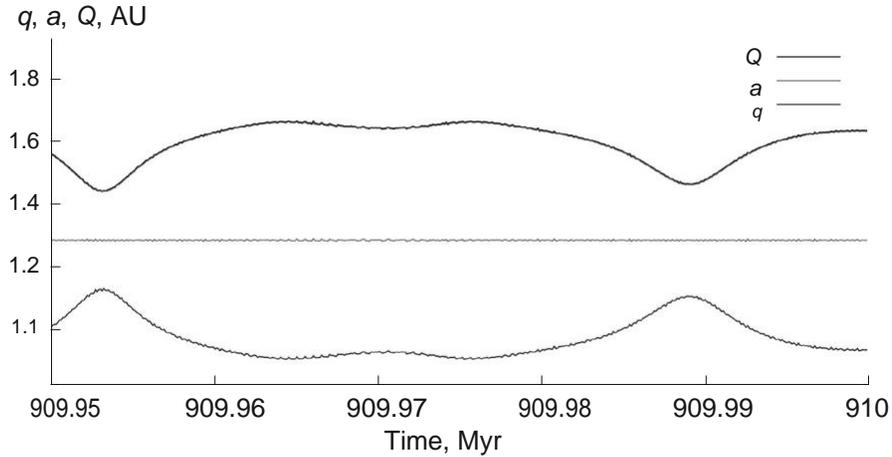

**Fig. 5.** Evolution with time (in Myr) of the semimajor axis *a*, perihelion and aphelion distances *q* and *Q* (in AU) of the orbits of the planetesimal at $a_0 = 1.28530$ AU, $a_{min} = 1.2$ AU, $e_0 = 0.15$, $k_C = 1$, and $t_S = 1^d$. The planetesimal moved in the 5:4 resonance with the planet Proxima Centauri *c*.

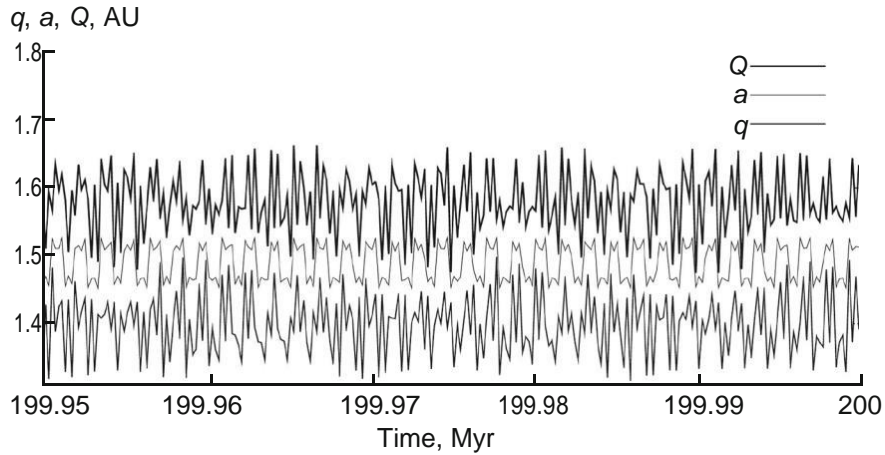

**Fig. 6.** Evolution with time (in Myr) of the semimajor axis *a*, perihelion and aphelion distances *q* and *Q* of orbits (in AU) of planetesimals at $a_0 = 1.50824$ AU, $a_{min} = 1.5$ AU, $e_0 = 0.02$, $k_C = 1$, and $t_S = 1^d$. The planetesimal moved in the resonance 1:1 with the planet Proxima Centauri *c*.

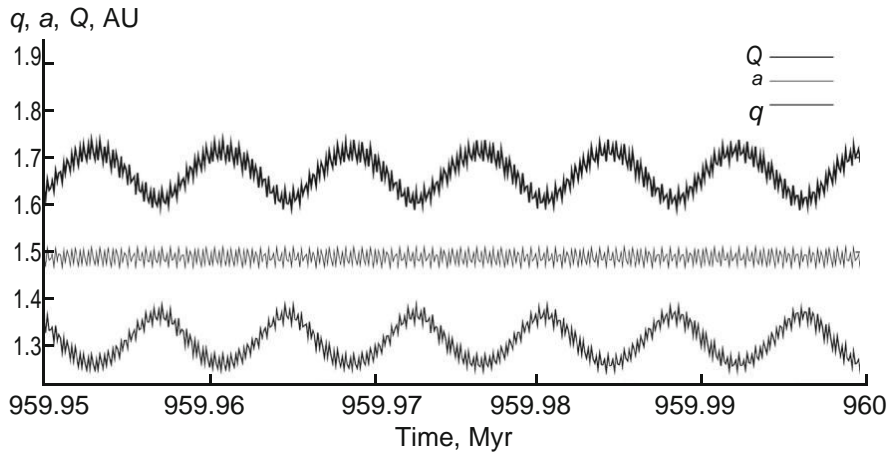

**Fig. 7.** Evolution with time (in Myr) of the semimajor axis *a*, perihelion and aphelion distances *q* and *Q* (in AU) of the orbits of the planetesimal at $a_0 = 1.50041$ AU, $a_{min} = 1.5$ AU, $e_0 = 0.15$, $k_C = 1$, and $t_S = 1^d$. The planetesimal moved in the resonance 1:1 with the planet Proxima Centauri *c*.

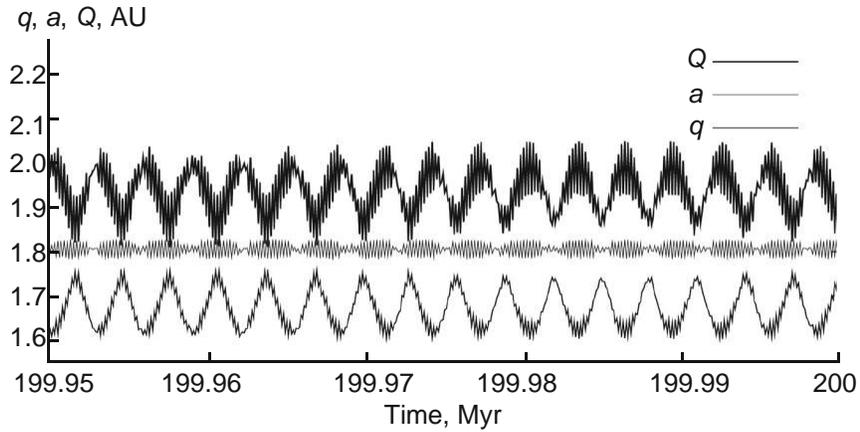

**Fig. 8.** Evolution with time (in Myr) of the semimajor axis *a*, perihelion and aphelion distances *q* and *Q* (in AU) of the orbits of the planetesimal at $a_0 = 1.78634$ AU, $a_{min} = 1.7$ AU, $e_0 = 0.02$, $k_c = 1$, and $t_s = 1^d$. The planetesimal moved in the resonance 3:4 with the planet Proxima Centauri *c*.

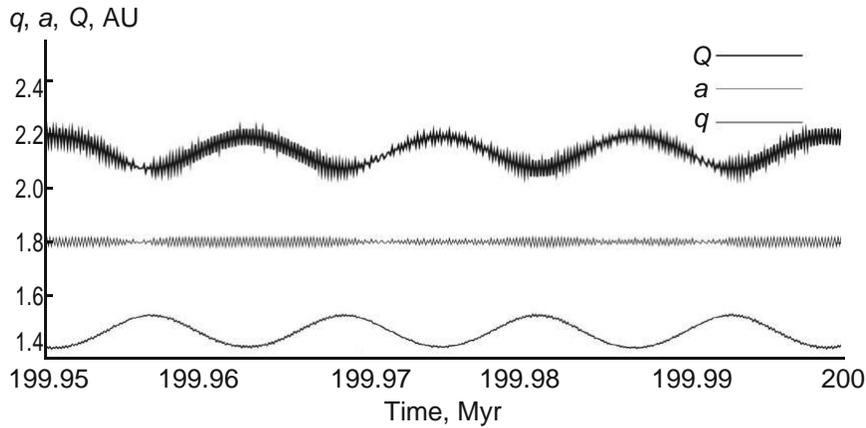

**Fig. 9.** Evolution with time (in Myr) of the semimajor axis *a*, perihelion and aphelion distances *q* and *Q* (in AU) of the orbits of the planetesimal at $a_0 = 1.78830$ AU, $a_{min} = 1.7$ AU, $e_0 = 0.15$, $k_c = 1$, and $t_s = 1^d$. The planetesimal moved in the 3:4 resonance with the planet Proxima Centauri *c*.

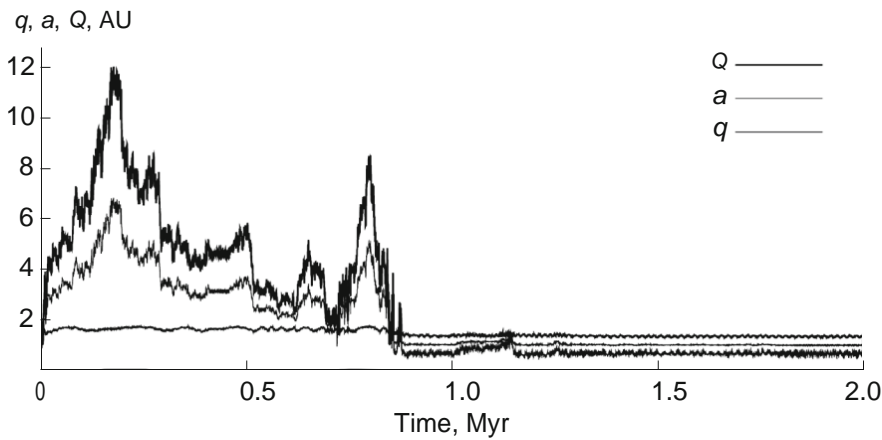

**Fig. 10.** Evolution with time (in Myr) of the semimajor axis *a*, perihelion and aphelion distances *q* and *Q* (in AU) of the orbits of the planetesimal at $a_0 = 1.53877$ AU, $a_{min} = 1.5$ AU, $e_0 = 0.15$, $k_c = 1$, and $t_s = 2^d$. The planetesimal moved in the 5:2 resonance with the planet Proxima Centauri *c* after 0.9 Myr.

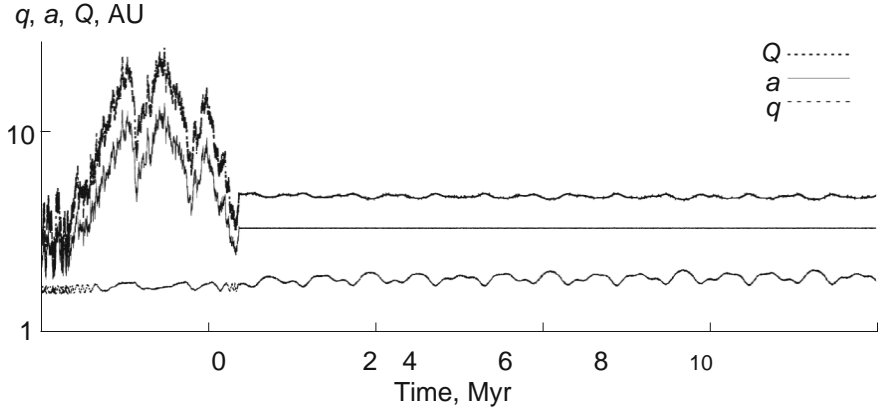

**Fig. 11.** Evolution with time (in Myr) of the semimajor axis $a$, perihelion and aphelion distances $q$ and $Q$ (in AU) of the orbits of the planetesimal at $a_0 = 2.01877$ AU, $a_{min} = 2.0$ AU, $e_0 = 0.15$, $k_C = 1$, and $t_S = 2^d$. The planetesimal moved in the resonance 3:10 with the planet Proxima Centauri $c$ after 2.4 Myr.

continued to move in elliptical orbits, was (1.478, 1.502 AU) at $k_C = 0.1$, there were two ranges (1.4805, 1.4848 AU) and (1.493, 1.498 AU) (with three values of $a_{i0}$ between them) $k_C = 0.5$, and there was only one value $a_{i0} = 1.504$ AU at $k_C = 1$. That the planetesimals moved in the resonance 1:1 with different masses of the planet may testify in favor of the fact that some of the Jupiter Trojans are not alien bodies, but those that could have remained since the formation of Jupiter.

## CONCLUSIONS

The feeding zone of the planet Proxima Centauri $c$ was considered, including subintervals of semimajor axes of stable orbits inside the main feeding zone and gaps outside the main zone. The research is based on the results of modeling the evolution of planetesimals' orbits under the inf luence of the star and planets Proxima $c$ and $b$ over a time interval of up to a billion years. Initial eccentricities $e_0$ of planetesimal orbits were 0.02 or 0.15, and their initial inclinations were $e_0/2$ radians.

If we consider the initial values of the product $a \times e$, which characterizes the change in the distance from the star to the moving planet Proxima $c$ and planetes-imals (for example, $e_c a_c$ and $e_0 a_{min002}$), then we have

$$a_c - a_{min002} = 0.04a_c + 0.02a_{min002} + 2.54\ a_c\mu^{1/3},$$

$$a_{max002} - a_c = 0.04a_c + 0.02a_{max\ 002} + 2.40a_c\mu^{1/3},$$

$$a_c - a_{min015} = 0.04a_c + 0.15a_{min015} + 2.23a_c\mu^{1/3}, \text{ and}$$

$$a_{max015} - a_c = 0.04a_c + 0.15a_{max015} + 4.3a_c\mu^{1/3},$$

Where $a_c$ and $e_c = 0.04$ are the semimajor axis and the eccentricity of the orbit of the planet Proxima $c$, $\mu$ is the ratio of the mass of the planet Proxima $c$ to the mass of the star, $a_{min002}$, $a_{max002}$, $a_{min015}$ and $a_{max015}$ are the minimum and maximum initial values of the semi-major axes of the considered planetesimals' orbits in the feeding zone of the planet Proxima $c$ at initial eccentricities $e_0$ of the orbits of planetesimals, equal to 0.02 and 0.15, respectively. Coefficients before $a_c\mu^{1/3}$ in the above three formulas are approximately 2.2– 2.5, i.e., are close to the coefficients in $\gamma = (2.1–2.45)\mu^{1/3}$ for circular initial orbits.

After the accumulation of the planet Proxima $c$, some planetesimals may have continued to move in stable elliptical orbits within its feeding zone, mostly cleared of planetesimals. Usually such planetesimals can move in some resonances with the planet Proxima $c$, for example, in the resonances 1:1 (as Jupiter Tro-jans), 5:4 and 3:4 and usually have small eccentrici-ties. Some planetesimals that moved for a long time (1–2 million years) along chaotic orbits fell into resonances the 5:2 and 3:10 with the planet Proxima $c$ and moved in them for at least tens of millions of years.

The dynamic characteristics of the planetary sys-tem near Proxima Centauri are such that there can be more analogs of the asteroid and trans-Neptunian belts in this system than in the Solar System. The smaller ratio of the mass of the planet Proxima $c$ to the mass of the star than that of Jupiter, the greater ratio of the semimajor axes of the orbits of the planets Proxima $c$ and $b$ than a similar ratio for Jupiter and Mars, and only one large planet in the Proxima Centauri system can be the reasons for such possible differences in the belts and the possible existence of a planet(s) between the orbits of the Proxima planets $b$ and $c$.

## ACKNOWLEDGMENTS

The author expresses his deep gratitude to the referees for useful remarks that contributed to the improvement of the paper.

## FUNDING

The research was supported by the grant 075-15-2020-780 of the Ministry of Science and Higher Education of the Russian Federation.